\documentclass[aps,pra,preprint,groupedaddress,showpacs]{revtex4}

\newcommand{\QED}{\hspace*{\fill}\mbox{\rule[0pt]{1.5ex}{1.5ex}}}
\def\tphi{\tilde{\phi}}

\def\openone{\leavevmode\hbox{\small1\kern-3.8pt\normalsize1}}
\def\RR{{\rm I\kern-.2emR}}

\def\tr{{\rm tr}\; }

\def\cb{{\cal B}}

\def\ch{{\cal H}}
\def\cs{{\cal S}}

\def\cp{{\cal P}}

\def\id{{\rm id}}

\newcommand{\ket}[1]{| #1 \rangle}
\newcommand{\bra}[1]{\langle #1 |}
\newcommand{\proj}[1]{\ket{#1}\! \bra{#1}}

\newcommand{\bitem}{\begin{itemize}}
\newcommand{\eitem}{\end{itemize}}
\newcommand{\benum}{\begin{enumerate}}
\newcommand{\eenum}{\end{enumerate}}
\newcommand{\beq}{\begin{equation}}
\newcommand{\eeq}{\end{equation}}
\newcommand{\beqa}{\begin{eqnarray}}
\newcommand{\eeqa}{\end{eqnarray}}
\newtheorem{definition}{Definition}

\newtheorem{theorem}{Theorem}
\newtheorem{example}{Example}
\newtheorem{proposition}{Proposition}

\newtheorem{lemma}{Lemma}
\newtheorem{corollary}{Corollary}
\newtheorem{conjecture}{Conjecture}

\newtheorem{remark}{Remark}
\newcommand{\bproof}{\begin{proof}}
\newcommand{\eproof}{\end{proof}}
\newcommand{\bprop}{\begin{proposition}}

\newcommand{\bdef}{\begin{definition}}

\def\C{{\bf C}}
\def\R{{\bf R}}

\begin{document}


\title{Better bound on the exponent of the radius of the multipartite separable ball}


\author{Leonid Gurvits and Howard Barnum}
\affiliation{ CCS-3, Mail Stop B256, Los Alamos National Laboratory,
Los Alamos, NM 87545} 
\affiliation{}


\date{May 20, 2005}

\begin{abstract}
We show that for an $m$-qubit quantum system, there is 
a ball of radius asymptotically approaching 
$\kappa  2^{-\gamma m }$ in Frobenius norm,
centered at the identity matrix,   
of separable (unentangled) positive semidefinite
matrices, for an exponent $\gamma = 0.5(\frac{\ln{3}}{\ln{2}} - 1) 
\approx .29248125$ much smaller in magnitude than the best 
previously known exponent, from our earlier work, of $1/2$.  
For normalized $m$-qubit states, we get a separable ball of 
radius $\sqrt{3^{m+1}/(3^m+3)} \times  2^{-(1 + \gamma)m} \equiv  
\sqrt{3^{m+1}/(3^m+3)}\times 6^{-m/2}$ (note that $\kappa = \sqrt{3}$), compared to the previous
$2 \times 2^{-3m/2}$. 
This implies that with parameters realistic for current experiments,
NMR with standard pseudopure-state preparation techniques
can access only unentangled states if 36 qubits or fewer
are used (compared to 23 qubits
via our earlier results).  We also obtain an improved exponent for
$m$-partite systems of fixed local dimension $d_0$, although approaching
our earlier exponent as $d_0 \rightarrow \infty$.   
\end{abstract}

\pacs{03.65.Ud,03.67.-a,03.67.Lx}

\maketitle

\newcommand{\thm}{\begin{theorem}}
\newcommand{\lem}{\begin{lemma}}
\newcommand{\pro}{\begin{proposition}}
\newcommand{\dfn}{\begin{definition}}
\newcommand{\rem}{\begin{remark}}
\newcommand{\xam}{\begin{example}}
\newcommand{\cnj}{\begin{conjecture}}
\newcommand{\que}{\begin{question}}
\newcommand{\cor}{\begin{corollary}}
\newcommand{\prf}{\noindent{\bf Proof:} }
\newcommand{\ethm}{\end{theorem}}
\newcommand{\elem}{\end{lemma}}
\newcommand{\epro}{\end{proposition}}
\newcommand{\edfn}{\bbox\end{definition}}
\newcommand{\erem}{\bbox\end{remark}}
\newcommand{\exam}{\bbox\end{example}}
\newcommand{\ecnj}{\bbox\end{conjecture}}
\newcommand{\eque}{\bbox\end{question}}
\newcommand{\ecor}{\end{corollary}}

\def\sup{^}
\def\Tp{Tchebyshef polynomial}
\def\Tps{Tchebyshef polynomials}
\newcommand{\rarrow}{\rightarrow}
\newcommand{\larrow}{\leftarrow}
\newcommand{\grad}{\bigtriangledown}

\overfullrule=0pt
\def\setof#1{\lbrace #1 \rbrace}
\section{Introduction and summary of results}

The existence of a ball of separable (that is, unentangled)
multipartite quantum states around the normalized identity matrix, and
estimates of the size of the largest such balls in various norms, are
important for a variety of reasons.  For example, lower estimates of
the sizes of balls provide easy to compute sufficient criteria for
separability of quantum states, as well as important tools for
studying the complexity of questions about entanglement and
multipartite quantum states.

A series of papers has established the existence \cite{Zyczkowski98a} and
provided successively better lower estimates \cite{Vidal99b, Braunstein99a,
Rungta2001a, GB02, GB03} of the sizes of these balls, notably of the
ball in $2$-norm (Frobenius norm).

In this paper, we use the same general idea we used in \cite{GB03} to
obtain the best previously known lower estimate: the idea of
considering the cone generated by tensor products of elements of the
cone generated by a ball of separable quantum states on 
some multipartite system and elements
of the cone generated by all quantum states on an additional single-party
system.  This cone
will consist of separable matrices by construction;  we find a lower
bound on the radius of a ball inside it, thereby providing a lower
estimate on the separable ball in the full system, though of smaller
radius than the separable ball we started with on one of the subystems.  
By inductively or recursively combining systems in this way, we obtain
lower estimates, dependent on the number of systems and their dimension,
of the size of the separable ball in a multipartite quantum system.

Here, we improve some aspects of our application of this technique, to
obtain a better lower estimate of the size of the ball in the convex
hull of the two cones (the ball-generated cone and the standard
separable cone on different systems).  When we apply the same
inductive strategy as in \cite{GB03}, we get a ball exponentially
larger in the number of combined systems.  For an $m$-partite quantum
with each subsystem having dimension $d_0$,  we get a ball of radius 
\beq
\left(\frac{d_0}{2d_0-1}\right)^{m/2-1}\;,
\eeq
in Frobenius
norm, centered at the identity matrix, of separable (unentangled)
positive semidefinite matrices (actually we do slightly better, but with the
same asymptotic exponent).
For qubits ($d_0 = 2$) this radius is 
is $(2/3)^{m/2-1}$, to be compared
to $(1/2)^{m/2-1}$ from \cite{GB03}.  If we express it as
as
$\kappa 2^{-\gamma m }$, the exponent is $\gamma =
0.5(\frac{\ln{3}}{\ln{2}} - 1) \approx .29248125$, compared
to \cite{GB03}'s exponent
of $\gamma=1/2$.  The non-qubit exponent is better, too, but
approaches our earlier one as $d_0 \rightarrow \infty$.
From this, we easily obtain a lower 
bound on the radius
of the largest Frobenius-norm ball of separable {\em normalized} density matrices:
for example, for $m$ qubits it is 
$ (3/2) \times 2^{-(1 + \gamma)m} \equiv 
(3/2) \times 6^{-m/2}$ (versus our earlier $2 \times 2^{-3m/2}$).
A slightly better, but more 
complicated, version of our new bound lets us improve the factor $3/2$ to 
$\sqrt{3^{m+1}/(3^m+3)}$, which rapidly approaches $\sqrt{3}$.  
This
gives a number of qubits
below which NMR with standard pseudopure-state preparation techniques
can access only unentangled states; with parameters realistic for
current experiments, this is 36 qubits (compared to 23 qubits via our
earlier results).  

We also address several points not strictly necessary for obtaining
these results, but which relate to the power and nature of our
methods, and the possibilities for strengthening the results.
Szarek \cite{Szarek2004a} found the first upper bound below unity 
on ball size, and recently Aubrun and Szarek \cite{Aubrun2005a} 
found an upper bound on ball 
size which  matches (up to a logarithmic factor) 
the lower bound we obtain here for qubits, though for qudits with $d>2$ there is 
still an exponential gap.  One of the 
most natural mathematical methods for tackling this problem is to use a
general result of F. John \cite{J48} relating the inner and outer ellipsoids of a 
convex set.  We show that straightforward application of
this natural method gives results weaker than we obtain here; 
weaker, in fact, than our earlier ones \cite{GB03}.  

Our methods may appear technical; nevertheless, many of the
intermediate results are mathematically interesting in their own right
and have applications to quantum information problems other than the
one at hand.  Along the way we explain some of these, notably a
variant proof of the result that the eigenvalues of a separable
bipartite quantum state are majorized by those of its marginal density
operators \cite{NK2001a}, and an example of the use of John's theorem to bound the
radii of other inner balls of quantum information-theoretic interest,
in this case the inner ball of the convex hull of all maximally
entangled states (related to an application-oriented entanglement
measure, the {\em fully entangled fraction} of \cite{GEJ02}).
Many of our results use bounds on induced norms of various classes
of maps on matrices, which we expect to be useful in other contexts.
An appendix includes an additional bound, closely related to one
used in the main argument, on the $2$-to-$\infty$ induced norm of stochastic
linear maps that are positive on a radius-$a$ ball of matrices around the
identity.

\section{Notation and mathematical preliminaries}

The basic definitions and notation we use, including many elementary
facts involving cones, positive linear maps, and duality, may be
found in \cite{GB03}.  Here we only review a few of the less-standard
of these.

We will use the term ``cone'' to mean a subset $K$ of a
finite-dimensional real vector space $V$ closed under multiplication
by positive scalars, which in addition we assume to be convex, pointed
(it contains no nonnull subspace of $V$) and closed in the
Euclidean metric topology.
The dual space of a real vector space
$V$ (the space of linear functions (``functionals'') 
from $V$ to $\R$) is written $V^*$.  The dual cone to
$C$ (the set of linear functionals which are nonnegative
on $C$) is $C^*$.
The adjoint of $\phi: V_1 \rightarrow V_2$ is 
$\phi^*: V_2^* \rightarrow V_1^*$, defined by
\beqa
\langle B, \phi(A) \rangle = \langle \phi^*(B), A \rangle \;,
\eeqa
for all $A \in V_1, B \in V_2^*$.  (Here we used $\langle B, A \rangle$
to mean the value of the linear functional $B$ evaluated on $A$.)
We say a linear map $\phi: V_1 \rightarrow V_2$ is $C_1$-to-$C_2$
positive, for cones $C_1 \subset V_1$, $C_2 \subset V_2$, if
$\phi(C_1) \subseteq C_2$.  When $C_2$ is a cone of positive semidefinite
(PSD) Hermitian matrices, we will sometimes abbreviate this to 
``$C_1$-positive.''

For complex matrices $M$, $M^\dagger$ denotes the transpose of the
entrywise complex conjugate of the matrix.  (The transpose itself is
$M^t$.)  $A \circ B$ denotes the elementwise (aka Hadamard or Schur)
product of two matrices, defined by $(A \circ B)_{ij} = (A)_{ij} (B)_{ij}$.
The positive semidefinite (PSD) cone in the real linear space
of Hermitian $d \times d$ matrices, is denoted $\cp(d)$.  We will
denote by ``$\succeq$'' the partial order induced by this cone ($X
\succeq Y$ iff $X - Y \in P(d)$); thus $M \succeq 0$ is
equivalent to $M \in \cp(d)$.  The linear space (over $\C$) of $N \times N$
complex matrices is denoted $M(N)$, and the linear space over the
reals of
$N \times N$ complex Hermitian matrices is denoted $\ch(N)$.  The
space of complex block matrices, $K$ blocks by $K$ blocks, with blocks
in $M(N)$, is denoted $\cb(K,N)$\\

Later, we will need the following easy proposition,
which follows from the fact 
that for normal (including
Hermitian) matrices,
$\Delta$, $||\Delta||_\infty$ is the largest 
modulus of an eigenvalue of $\Delta$.
\begin{proposition} \label{elementary, my dear Watson}
Let $\Delta$ be Hermitian.  If $||\Delta||_\infty \le 1$
then $I + \Delta \succeq 0$. 
\end{proposition}

We use the term  $m$-partite unnormalized density operator for
a positive semidefinite operator 
$$
\rho: H_{1} \otimes H_{2} \otimes ... \otimes H_{m} \longrightarrow
H_{1} \otimes H_{2} \otimes ... \otimes H_{m}
$$
We use the term $m$-partite unnormalized density matrix for a
matrix whose matrix elements
$$
\rho(i_{1},i_{2},...,i_{m}; j_{1},j_{2},...,j_{m}) 
$$
are those of an $m$-partite density operator in an orthonormal
basis constructed by choosing a fixed (ordered) orthonormal basis for
each subsystem, and taking all tensor products $e_{i_1} \otimes e_{i_2}
\otimes \cdots \otimes e_{i_m}$ of basis vectors for the subsystems.
We may view this as a block matrix partitioned according to the value
of, say, the first index; indeed, we may give it an $m$-level nested
block structure (given a choice of ordering of the indices).
Such a choice of local orthonormal bases and ordering of indices gives
an isomorphism between the space of operators on 
$H_1 \otimes \cdots \otimes H_m$ and a space of matrices (we may occasionally
implicitly identify these two spaces via an implicit isomorphism of this kind).   

\dfn
Consider cones $C_{i} \subset M(d_{i}),  1 \leq i \leq m$. A multipartite
unnormalized density matrix $\rho \in M(d_{1} d_{2} ... d_{m})$ 
(corresponding to an operator
$$
\rho: H_{1} \otimes H_{2} \otimes ... \otimes H_{m} \longrightarrow H_{1} \otimes H_{2} \otimes ... \otimes H_{m})
$$
is called $(C_{1} \otimes C_{2} \otimes ... \otimes C_{m})$-separable if it
belongs to the cone generated by the set $\{ A_{1}  \otimes A_{2} \otimes ... \otimes A_{m}: A_{i} \in C_{i} , 1 \leq i \leq m \}$.  We call this 
the separable cone, $\cs(C_1, C_2,..., C_m)$.
\end{definition}

This is trivially equivalent to the recursive definition:
$\cs(C_1, C_2,..., C_m)$ is the cone generated by the pairs 
$A \otimes B$ with $A \in S(C_2,...,C_{m-1}), B \in C_m$, 
and $\cs(C_1) := C_1$.

When  $C_i$ (for $1 \leq i \leq m$) are the PSD cones
$\cp(d_{i})$, 
$(C_{1} \otimes C_{2} \otimes ... \otimes
C_{m})$-separability is the standard notion of
separability of multiparty unnormalized density matrices.

We will use various norms on spaces of matrices or operators,
including the Frobenius or $2$-norm $||A||_2 := \sqrt{\tr A^\dagger A}$,
the $1$-norm $||A||_1 := \tr \sqrt{A^\dagger A}$, and the operator
norm $||A||_\infty := \max_{x: ||x||=1} ||Ax||$.  In the definition
of the operator norm, we used vector norms (written as $||\cdot||$)
on the input and output spaces, which we will take to be the Euclidean
norms induced by our chosen inner products on these spaces.  In 
general for linear operators $\phi: V \rightarrow W$ and norms
$||\cdot||_\tau$ and $||\cdot||_\omega$ on $V,W$ respectively, 
we will write
\beq
||\phi||_{\tau \rightarrow \omega} := 
\max_{x \in V: ||x||_\tau = 1} ||\phi(x)||_\omega\; ; 
\eeq
this is 
the operator norm induced by the choices $\tau,\omega$ for norms
on $V,W$.  Also, when $\phi: M(K) \rightarrow M(N)$ is Hermitian
preserving, we write $\phi^H$ for $\phi$'s restriction to Hermitian
matrices (i.e. to have
domain $\ch(K)$ and range $\ch(N)$).  These details are motivated
by the fact that key technical results of our paper involve
the relationship between  norms (induced by various choices
of matrix norms on the input and output matrix spaces) of 
linear maps $\phi: M(K) \rightarrow M(N)$, and similar norms
of $\phi^H$.  

Finally, a note on our usual choices for naming dimensions, which
should help make things clearer below.  When considering
a multipartite Hilbert space $H_1 \otimes H_2 \otimes \cdots \otimes H_m$, 
we use $d_1, d_2, ..., d_m$ for the dimensions of $H_1, ..., H_m$, 
and $d$ for the overall dimension $\Pi_{i=1}^m d_i$.
When we consider combining a ball cone and a PSD cone (as described
in the introduction and in more detail below), we let 
the ball cone be in a space of $d_2 \times d_2$ Hermitian matrices,
and the PSD cone in a space of $d_1 \times d_1$ Hermitian matrices.
When we consider linear maps between matrix spaces, we usually use
the somewhat unnatural choice that $M(d_2)$ (or $\ch(d_2)$) is 
the input space, and $M(d_1)$ ($\ch(d_1)$) the output space.  When
we consider an $m$-partite system where all the subsystems have
the same dimension, we use $d_0$ for the dimension of a local 
system and $d$ for the total dimension $d_0^m$.

\section{Main results}
\label{sec: main results}

We begin with some key definitions;  then we give an outline of the proof
of our main results, followed by the detailed proof.

\dfn 
If $X$ is a bipartite density matrix viewed as an
element of $\cb(d_1, d_2)$, so that its blocks
$X^{i,j}$ are in $M(d_{2})$, and if $\phi: M(d_{2}) \longrightarrow M(d_1)$ 
is a linear operator then we define 
\beq \tilde{\phi}(X) := \left(
\begin{array}{cccc} \phi(X^{1,1}) & \phi(X^{1,2}) & \dots &
\phi(X^{1,d_{1}})\\ \phi(X^{2,1}) & \phi(X^{2,2}) & \dots &
\phi(X^{2,d_{1}})\\ \dots &\dots & \dots & \dots \\ \phi(X^{d_{1},1})
& \phi(X^{d_{1},2}) & \dots & \phi(X^{d_{1} ,d_{1}})
\end{array} \right).
\eeq
\end{definition}

A simple result characterizing separability, 
but one fundamental to our argument, is:

\lem
\label{blockcrit}
Suppose that the cone $C(d_{2}) \subset \ch(d_{2}) \subset M(d_{2})$ .
Then $X$ is $\cp(d_{1}) \otimes C(d_{2})$-separable
iff $\tilde{\phi}(X) \succeq 0$ (i.e. is positive semidefinite) for all
stochastic $C(d_{2})$-positive linear operators $\phi: M(d_{2}) \longrightarrow
M(d_{1})$.  \end{lemma}

For the proof, see \cite{GB03}.

With these, we can sketch the proof of our main result, which applies
to a tensor product of systems of dimensions $d_1,d_2,...,d_n$.  It
is a recursion relation for a radius $a_n$ such that all matrices within
(or at) Frobenius norm distance $a_n$ of the identity are separable (i.e.
$P(d_1) \otimes P(d_2) \cdots P(d_n)$-separable):

\beq \label{the relation}
a_n \le a_{n-1} \sqrt{\frac{ d_n }{2(1 - a_{n-1}^2/(\Pi_{i=1}^{n-1} d_i))
(d_n -1) +1}} \;.
\eeq

\noindent
{\em Proof-outline:}

1.) Begin by letting $d_2$ in 
Lemma \ref{blockcrit} be the total dimension 
$\Pi_{i=1}^{n-1} d_i$ for our set of systems and $C(d_2)$ be the separable
(i.e. $P(d_1) \otimes \cdots \otimes P(d_{n-1})$-separable) cone for these
systems, and $d_1$ of the lemma correspond to $d_n$ for our $n$ systems,
so the lemma says $X$ is separable if and only if:
\beq \label{an equation}
\left(
\begin{array}{cccc} \phi(X^{1,1}) & \phi(X^{1,2}) & \dots &
\phi(X^{1,d_{1}})\\ \phi(X^{2,1}) & \phi(X^{2,2}) & \dots &
\phi(X^{2,d_{1}})\\ \dots &\dots & \dots & \dots \\ \phi(X^{d_{1},1})
& \phi(X^{d_{1},2}) & \dots & \phi(X^{d_{1} ,d_{1}})
\end{array} \right) \succeq 0
\eeq
when $\phi(\cs(d_2,...,d_n)) \subseteq P(d_1)$ and $\phi(I)=I$.

2.) Since the ball Ball$(a_{n-1})$ of radius $a_{n-1}$ around 
the identity is separable by hypothesis, the 
set of stochastic operators $\phi$ that are positive on that ball 
is no smaller than those positive on the
separable matrices, so $X$ is separable if (\ref{an equation}) holds
for all such $\phi$.  Let $X = I + 
Y$, $Y$ Hermitian and traceless;
by Proposition \ref{elementary, my dear Watson},  $X$ is separable if
\beq \label{another equation}
||\tilde{\phi}(Y)||_\infty \le 1\;. 
\eeq

3.) For stochastic $\phi$ with $\phi({\rm Ball}(a_{n-1})) \succeq 0$, we easily
show $||\phi(M)||_\infty \le (1/a_{n-1}) ||M||_2$ when $M$ is Hermitian, 
while for $M$ traceless but not necessarily Hermitian we obtain
\beq  
||\phi(M)||_\infty \le \lambda \equiv 
(1/a_{n-1})\sqrt{2(1 - a^2/d_1 d_2 \cdots d_{n-1})} ||M||_2\;. 
\eeq
4.) We bound the LHS of (\ref{another equation})
with elementary norm inequalities (for typographic clarity, inside the norm
delimiters, we omit the curved braces that otherwise delimit
block matrices):
\beqa \label{yet another equation}
\left|\left|
\begin{array}{cccc} \phi(Y^{1,1}) & \phi(Y^{1,2}) & \dots &
\phi(Y^{1,d_{1}})\\ \phi(Y^{2,1}) & \phi(Y^{2,2}) & \dots &
\phi(Y^{2,d_{1}})\\ \dots &\dots & \dots & \dots \\ \phi(Y^{d_{1},1})
& \phi(Y^{d_{1},2}) & \dots & \phi(Y^{d_{1} ,d_{1}})
\end{array} \right|\right|_\infty  \le \nonumber \\ 
\left|\left|
\begin{array}{cccc} ||\phi(Y^{1,1})||_\infty  & ||\phi(Y^{1,2})||_\infty & \dots &
||\phi(Y^{1,d_{1}})||_\infty \\ || \phi(Y^{2,1}) ||_\infty & 
|| \phi(Y^{2,2}) ||_\infty & \dots &
|| \phi(Y^{2,d_{1}}) ||_\infty \\ \dots &\dots & \dots & \dots \\ 
|| \phi(Y^{d_{1},1}) ||_\infty
& || \phi(Y^{d_{1},2}) ||_\infty & \dots & || \phi(Y^{d_{1} ,d_{1}}) ||_\infty
\end{array} \right|\right|_\infty 
\nonumber \\
\le 
\left|\left|
\begin{array}{cccc} a_{n-1}^{-1}||Y^{1,1}||_2 & \lambda ||Y^{1,2}||_2 & \dots &
\lambda ||Y^{1,d_{1}}||_2 \\ \lambda ||Y^{2,1}||_2 & a_{n-1}^{-1}||Y^{2,2}||_2 & 
\dots &
\lambda || Y^{2,d_{1}}||_2 \\ \dots &\dots & \dots & \dots \\ \lambda ||Y^{d_{1},1}||_2
& \lambda ||Y^{d_{1},2}||_2 & \dots & a_{n-1}^{-1} ||Y^{d_{1} ,d_{1}}||_2
\end{array} \right|\right|_\infty 
\;,
\eeqa
where we used the bounds from step 3.), along with the fact that $Y$'s offdiagonal
blocks may be made traceless by local transformations without affecting its 
separability or entanglement, in the last inequality.

5.) We prove an upper bound on $||\phi_B||_{2 \rightarrow \infty}$ for
maps $\phi_B: Z \mapsto B \circ Z$, and evaluate it
in the case that $B$'s matrix elements 
are equal to a constant on the diagonal, and another constant off the diagonal.
Calling this upper bound $\mu_B$, we have 
$||\phi_B(Z)||_\infty \le \mu_B ||Z||_2$;  we apply it to the last expression
in step 4.) to get: 
\beqa
\left|\left|
\begin{array}{cccc} a_{n-1}^{-1}||Y^{1,1}||_2 & \lambda ||Y^{1,2}||_2 & \dots &
\lambda ||Y^{1,d_{1}}||_2 \\ \lambda ||Y^{2,1}||_2 & a_{n-1}^{-1}||Y^{2,2}||_2 & 
\dots &
\lambda || Y^{2,d_{1}}||_2 \\ \dots &\dots & \dots & \dots \\ \lambda ||Y^{d_{1},1}||_2
& \lambda ||Y^{d_{1},2}||_2 & \dots & a_{n-1}^{-1} ||Y^{d_{1} ,d_{1}}||_2
\end{array} \right|\right|_\infty  
\nonumber \\
\le \mu_B
\left|\left|
\begin{array}{cccc} ||Y^{1,1}||_2 &  ||Y^{1,2}||_2 & \dots &
 ||Y^{1,d_{1}}||_2 \\  ||Y^{2,1}||_2 & ||Y^{2,2}||_2 & 
\dots &
 || Y^{2,d_{1}}||_2 \\ \dots &\dots & \dots & \dots \\  ||Y^{d_{1},1}||_2
&  ||Y^{d_{1},2}||_2 & \dots & ||Y^{d_{1} ,d_{1}}||_2
\end{array} \right|\right|_2 
\nonumber \\
= \mu_B ||Y||_2\;.
\eeqa

By step 2, then, $\mu_B ||Y||_2 \le 1$, i.e. $||Y||_2 \le \mu_B^{-1}$ 
implies separability of 
$X \equiv I + Y$.
We have 
\beq 
B = 
\left(
\begin{array}{cccc}  a_{n-1}^{-1} &  \lambda & \dots &
 \lambda \\ \lambda  & a_{n-1}^{-1} & 
\dots &
 \lambda \\ \dots &\dots & \dots & \dots \\ \lambda
&  \lambda & \dots & a_{n-1}^{-1}
\end{array} \right)\;,
\eeq
and we will show that $\mu_B$ works out to be 
\beq
\sqrt{\frac{\lambda^2 (d_n - 1) + a^2}{d_n}}\;.
\eeq
Using the expression we will derive for  $\lambda$ gives that
\beqa
||Y||_2 \le \sqrt{\frac{d_n}{\lambda^2 (d_n - 1) +1 }} \nonumber \\
= \sqrt{\frac{d_n}{2(1 - a^2_{n-1}/\Pi_{i=1}^{n-1} d_i) (d_n - 1) +1 }}
\eeqa
guarantees separability of $I + Y$, establishing
(\ref{the relation}).

We will apply our results also to balls of separable normalized states,
using the following result taken over from  
\cite{GB03} (where it is Proposition 7).
This  proposition is derived using
``scaling,''
i.e., considering all ways of writing a matrix
$\rho$ as a positive scalar times the sum of the identity and a
Hermitian perturbation, and minimizing the 2-norm of the 
perturbation. 
\begin{proposition} \label{jalopy}
Define $\mu(\rho)$ as the maximum of $||\Delta||_2$ over
all $\Delta$ such that there exists an $\alpha > 0$ for 
which $\rho = \alpha(I + \Delta)$.
Let $\rho$ be a normalized ($\tr \rho = 1$) density matrix.
Then the following three statements are equivalent:\\
1. $\mu(\rho) \le a$.\\
2. $\tr \rho^2 \le 1/(d-a^2)$.\\
3. $||\rho - I/d||_2 \le a/\sqrt{d(d-a^2)}$. 
\end{proposition}

\cor \label{corollary: general normalized}
Let $a$ be a lower bound on the size of the $m$-partite 
separable ball around
the identity matrix, $d$ be the dimension of the $m$-partite
Hilbert space.
If an $m$-partite normalized (i.e. unit trace) density matrix $\rho:
H_1 \otimes \cdots \otimes H_m 
\longrightarrow 
H_1 \otimes \cdots \otimes H_m$ satisifes $||\rho - I/d ||_2 
\le \frac{a}{d}$,
where $d = dim ( H_1 \otimes \cdots \otimes H_m )$, then it is separable.
\ecor
(The proposition actually gives the (negligibly) 
tighter statement with 
$\sqrt{d(d-a^2)}$ in the denominator.)

We now embark on a more detailed presentation and proof of our results,
beginning with some definitions.

\begin{definition}
Let $G(N,a) \subset \ch(N) \subset M(N)$ be the cone generated by
hermitian $N \times N$ matrices of the form $\{I +\Delta :
||\Delta||_{2} := (tr(\Delta\Delta^{\dagger})^{\frac{1}{2}} \leq a \}
$.
\end{definition}

Let $\phi: M(d_2) \rightarrow M(d_1)$ be stochastic.
Consider the maximum ``contraction or dilation ratio'' of
$\phi$ on Hermitian operators,   
\beq
\max_{{\rm Hermitian~} A}||{\phi}(A)||_\infty / ||A||_2\;.
\eeq
Note that this is equal to $\max_{ ||A||_2=1, A 
{\rm ~Hermitian}} ||{\phi}(A)||_\infty $, and therefore equal
to 
\beq 
||\phi^H||_{2 \rightarrow \infty}\;.
\eeq

\begin{definition}
Define $\gamma(d_1,d_2, a)$ as
the maximum, over stochastic maps 
$\phi : M(d_2) \rightarrow
M(d_1)$ that are positive on 
$G(d_2,a)$, of 
 $||\tilde{\phi}^H||_{2 \rightarrow \infty}$.  
\end{definition}
Note that we
used $\tilde{\phi}$ here, not $\phi$ itself.
\begin{proposition} 
\label{prop: induction step general}
Let $H_1, H_2$ have dimensions
$d_1, d_2$.  
If an unnormalized density matrix $\rho: H_{1} \otimes
H_{2}  \longrightarrow H_{1}  \otimes H_{2}$ 
satisfies the inequality $||\rho-I||_{2} \leq 1/\gamma(d_1,d_2,a)$ then
it is $\cp(d_1) \otimes G(d_2, a)$-separable.  
\end{proposition}

\noindent
{\em Proof: }
Let $\rho = I + \Delta$,  $\Delta$ Hermitian;  by Lemma $1$, 
we are looking for a bound on $||\Delta||_2$ that ensures, 
for any stochastic $G(d_2,a)$-positive linear operator
(i.e. $\phi(X) \succeq 0$ for all $X \in G(d_2,a)$), that
$\tilde{\phi}(I + \Delta) \succeq 0$.  $\tphi(I) = I$, so 
$\tphi(I + \Delta) = I + \tphi(\Delta)$;  
$||\tphi(\Delta)||_\infty \le 1$
will ensure $\tphi(I + \Delta) \succeq 0$ (cf. Proposition 
\ref{elementary, my dear Watson}).  Since 
$||\tphi(\Delta)||_\infty/||\Delta||_2 \le \gamma(d_1,d_2,a)$
from the definition of $\gamma(d_1,d_2,a)$, 
$||\Delta||_2 \le 1/\gamma(d_1,d_2,a)$ ensures this. \QED

In order to make good use of this proposition, we need a bound on the
value of $\gamma(d_1,d_2,a)$.  Proposition \ref{prop: gamma bound}
below, together with Proposition \ref{prop: lambda bound}'s bound on the
parameter $\lambda$ that appears in Proposition \ref{prop: gamma bound}, provides it.
Obtaining this bound on $\gamma$ 
is the technical heart of our results, and the
improvement in this bound over that found in \cite{GB03} is the source
of the better exponent in the lower bound on the size of the separable
ball we obtain in the present paper.  We begin with a definition and 
an easy lemma.

\begin{definition}
Define $\lambda(d_1,d_2, a)$ as the maximum, over all    
stochastic maps $\phi: M(d_2) \rightarrow M(d_1)$, 
positive on $G(d_2,a)$, and over all traceless $X \in M(d_2)$, of  
$||\phi(X)||_\infty/||X||_2$.
\end{definition}

\begin{lemma}
\label{lemma: contract}
 If $\phi: M(d_2) \longrightarrow M(d_1)$ is a stochastic $G(d_2,a)$-positive
linear map
with $0 \le a \le 1$, and
$\phi(I) = I \in M(d_1)$, then
\\ \noindent
$||\phi(X)||_{\infty} \leq a^{-1} ||X||_{2} $ for all $X \in \ch(d_2)$.
\end{lemma}

\noindent
{\bf Proof:}
$G(d_2,a)$-positivity of a stochastic $\phi$
means $||\phi(\Delta)||_\infty \le 1$ for all Hermitian
$||\Delta||$ with $||\Delta||_2 \le a$;  since $||\phi(\Delta)||_\infty$
is homogeneous in $||\Delta||_2$, it will achieve its maximum on such $\Delta$
where $||\Delta||_2= a$, implying $||\Phi(\Delta)||_\infty / ||\Delta||_2
\le 1/a$.
\QED

We now proceed to our key bound, on $\gamma(d_1, d_2, a)$.  

\begin{proposition} \label{prop: gamma bound}
Suppose $a > 1/d_2$.  Then
\beqa
\gamma(d_1,d_2,a) 
\le  a^{-1}\sqrt{\frac{a^2\lambda^2(d_1,d_2, a)(d_1-1) + 1}{d_1}}\;.
\eeqa
\end{proposition}

\noindent
{\em Proof:}
Let $A \in \cb(d_1,d_2)$ be a Hermitian 
$d_1 \times d_1$ matrix of $d_2 \times d_2$ blocks
$A^{(i,j)}$.  Call the $2$-norms of the blocks $a^2_{ij}$, and the 
operator norms of the blocks $a^\infty_{i,j}$, and define $A^2$ and 
$A^\infty$ as the matrices with these elements.  Similarly, 
call the matrices 
whose elements are 
$||\phi(A^{(i,j)})||_{\{2,\infty\}}$, $\Phi^{\{2,\infty\}}$.  
(We promise not to 
square any matrices named $A$ or $\Phi$, 
so this notation is unambiguous.)

Note that $||A^2||_2 = ||A||_2$.  Also, note that 
\beq
||\tilde{\phi}(A)||_\infty \le ||\Phi^\infty||_\infty\;,
\eeq
by an elementary norm inequality (the operator norm of a block matrix 
is bounded above by the operator norm of the matrix whose elements
are the operator norms of the blocks of the original matrix). 
$\Phi_\infty$ is a matrix with nonnegative entries.  Its diagonal 
entries $||\phi(A^{(i,i)})||_\infty$ are bounded above by $a^{-1}||A^{(i,i)}||_2$
by Proposition \ref{lemma: contract}, which applies since the diagonal blocks
of $A$ are Hermitian.  The offdiagonal blocks are not in general Hermitian, 
but they may be made traceless via a unitary ``local transformation'' (acting only 
on the index specifying which block) which has 
no effect on the matrix's separability or entanglement.
This is because one of its (unnormalized) ``reduced density matrices,''
is the matrix of traces of its blocks, and  
the reduced matrix may be diagonalized by a local transformation.

So for the offdiagonal entries $\Phi^\infty_{i,j}$, $i \ne j$ we have 
$\Phi^\infty_{i,j} \equiv ||\phi(A^{(i,j)})||_\infty \le 
\lambda(d_1,d_2,a) ||A^{(i,j)}||_2$ by the definition of $\lambda$.  
In other words, using 
$\le$ for the ordering in which $A \le B$ means $B-A$ is (entrywise) nonnegative,
we have
\beqa
\Phi^\infty \le a^{-1} L  \circ  A^{(2)}\;,
\eeqa 
where $L$ is the matrix with $1$'s on the diagonal 
and $a \lambda(d_2, d_1, a)=: \eta$ in all
offdiagonal places.  Therefore (since the operator 
norm is monotonic in the ordering $\le$), the maximal contraction/
dilation ratio on Hermitian matrices, i.e. the 
$2$-to-$\infty$ induced norm (on Hermitian matrices), of
the completely positive map 
$\Lambda_a: X \mapsto a^{-1} L  \circ  X$
taking $M(K) \rightarrow M(K)$  
is an upper bound on $||\tilde{\phi}^H||_{2 \rightarrow \infty}$.  
The induced norm of $\Lambda_a$ is $a^{-1} ||\Lambda||_{2 \rightarrow \infty}$,
where $\Lambda: X \mapsto L \circ X$; we evaluate it via the
following Lemma.

\begin{lemma}\label{lemma: Schur contraction bound}
Let $\phi_B$ be the linear map from 
$\ch(n)$ to $\ch(n)$ defined by 
$\phi_B: X \mapsto B \circ X$, for some Hermitian $B$.  Then 
\beqa
||\phi_B||_{2 \rightarrow \infty} = 
\max_{y_i \ge 0, \sum_i y_i = 1; i \in \{1,..,n\}} 
y^t C y\;,
\eeqa
where  C is the $n \times n$ matrix with elements
$C_{ij} = |B_{ij}|^2$.
\end{lemma}

The lemma states that the 
$2$-norm-to-$\infty$-norm induced norm of the positive
map defined by the Schur (elementwise) 
product with $B$ for some fixed Hermitian $B$, 
is just the maximum value of a quadratic form over a simplex,
the matrix of the quadratic form being the one whose elements are
the absolute squares of $B$'s.
This lemma has independent interest; we defer its proof and a 
discussion of other applications to Section \ref{sec: schur map lemma}.

Recall the abbreviation
$a \lambda(d_2, d_1, a) =: \eta$, and note that the premise of
the Proposition we are proving implies $\eta > 1$.  We have
\beqa
||\Lambda||_{2 \rightarrow \infty}^2 = 
\max_{y \in \R_+^{d_1}, \sum_i y_i = 1}
(1 - \eta^2) \sum_i y_i^2 + \eta^2( \sum_{i} y_i )^2 \; \nonumber \\
 =  (1 - \eta^2) \sum_i y_i^2 + \eta^2 \;,
\eeqa
where we used $\sum_i y_i = 1$.  
Since $\eta \ge 1$, this is maximized where $\sum_i y_i^2$ is
minimized, i.e. with each $y_i = 1/d_1$.  The maximal value is
$(1 - \eta^2)/d_1 + \eta^2$, and thus 
\beq
||\Lambda||_{2 \rightarrow \infty} = 
\sqrt{\frac{\eta^2(d_1-1) + 1}{d_1}} \equiv
\sqrt{\frac{a^2\lambda^2(d_1-1) +1}{d_1}}\;.
\eeq

Since (as argued before Lemma \ref{lemma: Schur contraction bound})
$a^{-1}||\Lambda_2||_{2 \rightarrow \infty}$
is an upper bound on $||\tilde{\phi}^H||_{2 \rightarrow \infty}$, this
gives the desired result. \QED

\noindent
{\bf Remark:}  The ease with which we were able to use Lemma 
\ref{lemma: Schur contraction bound} in the above proof was
due to the simple form of the matrix $L$ which took the role
of $B$.  The problem of maximizing a general quadratic form with 
nonnegative matrix, over the simplex, is NP-hard as one 
can reduce Max-Clique to it (this is apparently
well-known, cf. \cite{BomzeDeKlerk2001a} or \cite{Bomze98a}).

To make further use of this in evaluating $\gamma(d_1, d_2, a)$, we need
an estimate
for $\lambda(d_2, d_1, a)$.  The following proposition provides one.

\begin{proposition} \label{prop: lambda bound}
\beq
\lambda(d_2, d_1, a) \le \frac{1}{a}\sqrt{2(1 - \frac{a^2}{d_2})}\;.
\eeq
\end{proposition}

This plays the role that Proposition 6 did in \cite{GB03}, 
but while that proposition did not assume $\phi$ stochastic,
and established that for all $\phi$ whose $2$-to-$\infty$-induced norm
on Hermitian operators is at most $1$,
the induced norm on  all operators
is at most $\sqrt{2}$, 
the present proposition adds the assumption of stochasticity, and
computes the maximum induced norm for the class of stochastic 
$G(d_2,a)$-positive
maps acting on traceless matrices, rather than all matrices.  
In fact, using Proposition 6 of \cite{GB03} for the
bound on $\lambda$ and the rest of the argument as in the present paper, 
we could have obtained the same exponent in 
our bound on ball size as a function of number of systems
$m$.

\noindent
{\bf Proof:}  
We need good bounds on  
the $2$-to-$\infty$ induced norms of $G(d_2,a)$-positive 
maps $\phi: M(d_2) \rightarrow M(d_1)$.  
Since it will turn out that these do not depend on 
$d_1$, we will use $d$ in place of $d_2$ throughout 
the discussion.    We consider normalized
matrices in $G(d, a)$, which are expressible as $\rho = I/d  + \Delta$ for 
some traceless Hermitian perturbation $\Delta$, and recall from Proposition 
\ref{jalopy} that these are precisely those normalized $\rho$ for
which $||\Delta||_2 \equiv ||\rho - I/d ||_2 \le 
a/\sqrt{d (d -a^2)}$.  $G(d,a)$-positivity is equivalent to positivity 
on these normalized matrices  
(since they generate the cone $G(d,a)$ by positive scalar multiplication).  
The latter is equivalent to the condition
\beqa
\phi(I/d  + \Delta) \succeq 0 {\rm~whenever~} ||\Delta||_2 \le 
a/\sqrt{d (d -a^2)}\;.
\eeqa
Using Proposition 1, for stochastic $\phi$ this is equivalent to 
\beqa \label{schmoozer}
||\Delta||_\infty \le 1/d  {\rm~whenever~} ||\Delta||_2 \le 
a/\sqrt{d (d -a^2)}\;.
\eeqa
For Hermitian traceless
$\Delta$, $||\phi(\Delta)||_\infty / ||\Delta||_2$ is homogeneous
of degree zero in $\Delta$, and therefore\beqa \label{schlemiel}
||\phi(\Delta)||_\infty / ||\Delta||_2 \le 
(1/d )/(a/\sqrt{d (d -a^2)}) \equiv  \nonumber \\ \label{groovy}
(1/a) \sqrt{1 - a^2/d }\;.
\eeqa
To extend this to arbitrary, not necessarily Hermitian, traceless
matrices $B$ write 
$B$ in terms of traceless Hermitian and traceless antiHermitian parts
as $B =  X + iY$.  Then
\beqa \label{schnuck}
||\phi(B)||_\infty & \le &  ||\phi(X)||_\infty + ||\Phi(Y)||_\infty
\nonumber \\
& \le & a^{-1}\sqrt{1-a^2/d } (||X||_2 + ||Y||_2) \nonumber \\
& \le &  a^{-1}\sqrt{1-a^2/d } \sqrt{2} b\;, \label{RHS}
\eeqa
where the second inequality is  (\ref{groovy}) and the last is
elementary Euclidean geometry.  Thus 
\beq
||\phi(B)||_\infty / ||B||_2 \le 
a^{-1}\sqrt{2(1-a^2/d)} \;.
\eeq
\QED

Incorporating the upper bound of Proposition \ref{prop: lambda bound}
explicitly into Proposition \ref{prop: gamma bound} gives

\begin{proposition} \label{prop: better second gamma bound}
\beqa
\gamma(d_1,d_2,a):= 
\max_{\phi} 
\max_{{\rm Hermitian~} A}||\tilde{\phi}(A)||_\infty / ||A||_2 \nonumber \\
 =  a^{-1}\sqrt{  \frac{2(1 -a^2/d_2)(d_1-1) + 1}{d_1}}\;.
\eeqa
\end{proposition}

Using this 
bound in Proposition \ref{prop: induction step general} gives:

\begin{proposition}

\label{prop: better induction step}
Let $H_1, H_2$ have dimensions
$d_1, d_2$.  
If an unnormalized density matrix $\rho: H_{1} \otimes
H_{2}  \longrightarrow H_{1}  \otimes H_{2}$ 
satisfies the inequality 
\beq
||\rho-I||_{2} \leq a \sqrt{\frac{d_1}{2(1 - a^2/d_2)(d_1-1) +1}}
\eeq
 then
it is $\cp(d_1) \otimes G(d_2, a)$-separable.  
\end{proposition}

We may apply this proposition inductively or recursively, in various
ways, to obtain bounds on multipartite separability.  In the following
the induction proceeds as in \cite{GB03}, by tensoring one additional
PSD cone $P(d_n)$ with a cone generated by a ball of separable states
in $P(d_1) \otimes \cdots \otimes P(d_{n-1})$, of radius $a_{n-1}$,
obtained in the previous inductive step.  The induction begins with
the base case of a bipartite separable ball of radius one (in 2-norm)
around the identity (from \cite{GB02}).
From Proposition 
\ref{prop: better induction step} 
we have the recursion relation:
\beq \label{eq: better recursion relation}
a_n \le a_{n-1} \sqrt{\frac{ d_n }{2(1 - a_{n-1}^2/(\Pi_{i=1}^{n-1} d_i))
(d_n -1) +1}} \;.
\eeq

This allows for easy numerical calculation of $a_n$.
When we have a total of $m$ systems each of dimension $d_0$, we have:
\beq \label{eq: better all the same dim}
a_n \le a_{n-1} \sqrt{\frac{ d_0 }{2(1 - a_{n-1}^2/d_0^{n-1})
(d_0 - 1) +1 }} \;.
\eeq

For qubits, this is 
\beq 
\label{eq: better qubits}
a_n \le a_{n-1} \sqrt{\frac{ 2}{3 - a_{n-1}^2 /2^{n-1}}} \;.
\eeq


Using the weaker bound
$\lambda(d_1,d_2,a) \le a^{-1}\sqrt{2}$ from \cite{GB03} gives a 
weaker but easily solved recursion relation:
\begin{proposition}
\label{prop: weaker induction step}
Let $H_1, H_2$ have dimensions
$d_1, d_2$.  
If an unnormalized density matrix $\rho: H_{1} \otimes
H_{2}  \longrightarrow H_{1}  \otimes H_{2}$ 
satisfies the inequality 
\beq
||\rho-I||_{2} \leq a \sqrt{\frac{d_1}{2(d_1-1) +1}}
\eeq
 then
it is $\cp(d_1) \otimes G(d_2, a)$-separable.  
\end{proposition}
This gives a worse bound, 
but asymptotically the same exponent for the number of 
systems:  
\begin{corollary}
\label{corollary: d-dimensional systems}
If an $m$-partite unnormalized density matrix $\rho:
H_1 \otimes \cdots \otimes H_m 
\longrightarrow 
H_1 \otimes \cdots \otimes H_m$ satisfies 
\beq
||\rho - I ||_2 
\le 
\left(\frac{d_0}{2d_0-1}\right)^{m/2 -1}
\eeq 
then it is separable.
\end{corollary}
While for large $d_0$, Corollary 
\ref{corollary: d-dimensional systems} is asymptotically the same as the
bound $(1/2)^{m/2-1}$ 
from \cite{GB03}, for qubits it gives the notably better
\begin{corollary}
\label{corollary: qubits}
If an $m$-qubit unnormalized density matrix $\rho:
H_1 \otimes \cdots \otimes H_m 
\longrightarrow 
H_1 \otimes \cdots \otimes H_m$ satisifes 
\beq
||\rho - I ||_2 
\le 
(2/3)^{m/2 - 1}
\eeq 
then it is separable.
\end{corollary}

In fact, we may explicitly solve the recursion (\ref{eq: better all the same dim})
exactly, obtaining:
\begin{theorem}
If an $m$-qudit unnormalized density matrix satisfies the inequality
\beqa
||\rho - I||_2 \le r_n :=  \sqrt{\frac{d^n}{(2d-1)^{n-2}(d^2-1) +1}}\;.
\eeqa
then it is separable.
\end{theorem}
For qubits, we have:
\beqa
r_n = \sqrt{\frac{2^n}{3^{n-1} +1}} \equiv \sqrt{\frac{3^{n+1}}{3^n + 3}}
2^{-\gamma m}
\eeqa
with $\gamma =
0.5({\frac{\ln{3}}{\ln{2}} - 1}) \approx .29248125$,
compared to $\gamma = 1/2$ from \cite{GB03}.

In an earlier version of the present paper we obtained the same exponent,
but a slightly worse overall expression, because we did not exploit local
transformations to render the offdiagonal blocks of $A$ in the proof 
of Proposition \ref{prop: gamma bound} traceless, and
so had to use a slightly worse contraction bound $\lambda'(d_1,d_2,a)$ that 
applies to all matrices, not just traceless ones.  Since this bound $\lambda'$
may prove useful in other situations, we include it and its proof in an 
appendix.  Subsequently,
Roland Hildebrand \cite{Hildebrand2005a} 
obtained the same asymptotic exponent but a slightly larger ball for the
$m$-qubit case, via an argument exploiting
the fact, special to the case of $m$ qubits, that
the local cones are already ball-generated (aka Lorentz) cones.  
In the proof above, we exploited the ability to render the offdiagonal 
blocks of $A$ traceless by local transformations, improving the
bound to agree with Hildebrand's 
in the qubit case, but also improving it for the case 
of $m$ $d_0$-dimensional systems (and indeed, in general).

Although Corollary \ref{corollary: qubits} gives a ball with $\kappa = 3/2$, as
mentioned above, we see that the present paper's 
improved bound on $\lambda(d_1, d_2, a)$, as embodied in 
(\ref{eq: better all the same dim}) gives a larger ball, with a prefactor 
$\kappa(m)$  asymptotically approaching $\sqrt{3}$.
For tripartite separability of unnormalized states,
Proposition \ref{prop: better induction step} gives a ball of radius
$\sqrt{4/5}$ around the identity (a result also noted by Hildebrand), 
larger than our previous result of $\sqrt{1/2}$.

Using  Corollary \ref{corollary: qubits}
and Proposition \ref{jalopy}, 
for $m$ qubits, we obtain a lower bound on the radius of
the largest {\em normalized} separable ball of $2^{-m}(2/3)^{m/2-1}$, 
i.e. $(3/2)\times 6^{-m/2} \equiv (3/2) \times 2^{-\eta m}$ with 
$\eta = (1/2) (\ln{6}/\ln{2}) \approx 1.292481$.
Using the stronger recursion we get 
$\sqrt{\frac{3^{n+1}}{3^n + 3}}
2^{-(1 + \gamma m)} \equiv \sqrt{\frac{3^{n+1}}{3^n + 3}}
6^{-m/2}$.
In  the course of investigating
the volume of the separable states relative to all normalized
states, Szarek (\cite{Szarek2004a} Appendix H) 
obtained
a lower bound of $6^{-m/2}$ on the radius of a related, but 
larger ``symmetrized'' set $\Sigma$, the convex hull of
$\cs \cup - \cs$).
In general case such symmetrization can substantially increase
the inner radius.  Indeed, in the case of the $d$-dimensional simplex, 
the inner ball has radius of
order 
$1/d$ compared to $1/\sqrt{d}$ for its symmetrization (which is the unit sphere
in $l_1$-norm).  
Szarek also
obtained the first {\em upper} bound below $o(2^{-m})$ on the radius of  balls
inside the normalized separable $m$-qubit states:  it is $o(2^{-\eta m})$  with 
the exponent $\eta  = 1 + (1/8)\log_2{27/16} \approx 1.094361$). 
Recently Aubrun and Szarek \cite{Aubrun2005a}
improved this, obtaining an upper bound for for the symmetrized
set of separable normalized states of qubits (which contains the separable states) 
of:
\begin{equation}
C_0 \sqrt{m \log{m}} 6^{-m/2}\;.
\end{equation}
The constant $C_0$ is equal to $\sqrt{3} C_1$, where $C_1$ (which appears
in a crucial lemma of \cite{Aubrun2005a}) can be chosen
to be $1.67263$, and can probably be chosen smaller. 
The asymptotic exponent for this expression matches 
that in our lower bound 
for the case of qubits, though with
the logarithmic prefactor.
On the other hand for $d \geq 3$  the inner radii in the unsymmetrized
and symmetrized cases are of different order.
Indeed, it is easy to prove that that the 
unnormalized separable radii
$r(d_1,...,d_k) \geq r(D_1,...,D_k)$ if $d_i \le  D_i ; 1 \leq i \leq k$.
In \cite{Aubrun2005a}, Aubrun and Szarek also state an upper bound 
of $(d_0(d_0 + 1))^{-m/2}$ (up to a similar prefactor)
for the normalized symmetrized qudit case, corresponding to 
order $((d+1)/d)^{-m/2}$ for the unnormalized ball around $I$.  This should
be compared to our results for the unnormalized ball
which are of order $((2d_0-1)/d_0)^{-m/2}$.  While
both of these give $(3/2)^{-m/2}$ in the case of qubits, the Aubrun-Szarek
exponent (with a constant base such as $2$) approaches zero as $d_0$ grows,
while ours does not (approaching, instead, $-1/2$).     
Thus in the case of $d_0 \ge 3$ there is still an
gap between our result and their upper bound, and it is an interesting
open problem to close this gap.
Notice that it had been proved in \cite{GB03} that the 
radius of the maximum  ball
inside the normalized real-separable $m$-qubit states is  $O(2^{-m})
\equiv O(1/d)$
(indeed, it is exactly $1/\sqrt{d(d-1)} = O(1/d)$ for general 
real-separable multipartite states).
We also showed in \cite{GB02} that the bipartite
separable states have in-radius $1/\sqrt{d(d-1)}$
(resolving a question raised, for example, in \cite{Kus2001a},
where the $d = 4$ case was proved).  
The $O(1/d)$ results correspond to 
a ball of radius order unity of unnormalized
real-separable or bipartite separable states,
compared with one that (from Szarek's upper bound) must
shrink as an inverse of a power of
dimension in the general unnormalized
multipartite case.   
This provides 
another example of a dramatic difference in the behavior of entanglement
in the bipartite versus the
multipartite situation.

\section{A $2$-to-$\infty$ induced norm bound for Schur product maps}
\label{sec: schur map lemma}

In this section, we prove Lemma \ref{lemma: Schur contraction bound}, 
which was used in proving Proposition
\ref{prop: gamma bound} in Section \ref{sec: main results}.  It states
that the maximum 
2-norm-to-$\infty$-norm contraction/dilation ratio for the positive
map defined by the Schur (elementwise) 
product with $B$ for some fixed Hermitian $B$, 
is just the maximum value of a quadratic form over a simplex,
the matrix of the quadratic form being the one whose elements are
the absolute squares of $B$'s.

\setcounter{lemma}{2}
\begin{lemma}
Let $\phi_B$ be the linear map from 
$\ch(n)$ to $\ch(n)$ defined by 
$\phi_B: X \mapsto B \circ X$, for some Hermitian $B$.  Then 
\beqa
||\phi_B||_{2 \rightarrow \infty} = 
\max_{y_i \ge 0, \sum_i y_i = 1; i \in \{1,..,n\}} 
y^t C y\;,
\eeqa
where  C is the $n \times n$ matrix with elements
$C_{ij} = |B_{ij}|^2$.
\end{lemma}

{\bf Proof:}
To show this, we use the basic fact (see e.g. \cite{Horn85a})
that for vector spaces (finite-dimensional, for simplicity)
$V,W$ equipped with norms $|| \cdot ||_V$, $|| \cdot ||_W$, 
and using the notation $||\cdot||_{V^*}, ||\cdot||_{W^*}$ for the norms
dual to $||\cdot||_V$, $||\cdot||_W$, for a fixed linear
map $T: V \mapsto W$
\beqa
\max_{X \in V} ||T(X)||_W / ||X||_V = 
\max_{Y \in W^*} ||T^*(Y)||_{V^*} / ||Y||_{W^*} \;.
\eeqa
Using the facts that
$\phi_B$ is its own dual ($\phi_B = \phi_B^*$), the $2$-norm is its own  
dual norm, and the operator norm is dual to the $1$-norm, 
we obtain that $\max_{||X||_2 = 1} ||\phi_B(X)||_\infty = \max_{||Y||_1 = 1} 
||\phi_B(Y)||_2$.  We proceed to evaluate the latter.

Since $||\phi_B(Y)||_2$ is increasing in $||Y||_1$, 
the maximization can be extended to the convex set  
$\{Y: ||Y||_1 \le 1\}$, and since $||\phi_B(Y)||_2$ is convex
the maximum will occur at an extremal
point of that set.  The extremal points of the ball of $n \times n$ 
Hermitian
matrices with $1$-norm at most $1$ are the rank-one projectors
(pure states) $X$  whose matrix elements are $x_i x^*_j$, for some 
normalized ($\sum_i x_i^2 = 1$) vector $x$. For such $X$, 
\beqa
(\phi_B(X))_{ij}= B_{ij} x_i x_j^* 
\;.
\eeqa
Hence 
\beqa
||\phi_B(X)||_2^2  =  \sum_i |B_{ii}|^2 |x_i|^4 + 
\sum_{i \ne j} B_{ij} B_{ji} |x_i|^2 |x_j|^2 \nonumber \\
 = \sum_i |B_{ii}|^2 |x_i|^4 + 
\sum_{i \ne j} |B_{ij}|^2 |x_i|^2 |x_j|^2 \;.
\eeqa
Defining $y$ as the vector in $\R_n^+$ with 
$y_i := |x_i|^2$ and the matrix $C$ by $C_{ij} = |B_{ij}|^2$,
this expression is just $y^t C y$, and we are to maximize it 
over $y \in \R^n_+$ such that $\sum_i y_i = 1$, establishing
the lemma. \QED

\noindent
{\bf Digression:}  
Completely positive maps of the form considered in Lemma 
\ref{lemma: Schur contraction bound} are useful in a variety
of contexts in quantum information theory.  Simplest, perhaps,
is their appearance in
the most general representation of ``partial decoherence''
processes in some basis.  The relevant mathematical fact here
is that the set of completely positive maps $T$ 
such that there exists an orthonormal basis 
$e_i$ for which $T(e_i e_i^\dagger) = e_i e_i^\dagger$, 
or equivalently all states diagonal in that basis are fixed points
of the map, is precisely the set of maps $X \mapsto B \circ X$ with
$B$ Hermitian and having ones on the diagonal.  These maps are
doubly stochastic, implying that the output density matrix is 
``more disordered'' than the input density matrix, meaning 
its eigenvalues are majorized by those of the input density matrix
(\cite{Ando89a}, 
Theorem 7.1).

Another
application is an alternative proof of a fact due
to Nielsen and Kempe \cite{NK2001a}, that the vector of decreasingly
ordered eigenvalues of a separable bipartite mixed state is majorized
by that 
of either of its marginals (reduced states):  ``separable states 
are more disordered globally than locally.''  The proof uses
the well-known fact, useful in a variety of contexts both within
and outside of quantum information, that the (necessarily PSD)
matrices $AA^\dagger$ and $A^\dagger A$
have the same eigenvalues.  Equivalently, 
a quantum state (even an unnormalized one)
\beq \label{eq: decomposition of state}
R = \sum_i v^i v^{i\dagger}
\eeq
has the same eigenvalues as the Gram 
matrix of the (not necessarily normalized!)
vectors $v^i$ (the matrix whose $ij$ element is
the inner product $\langle v^i, v^j \rangle 
\equiv v^{i\dagger} v^j$), as one sees by letting
$A$ in the above fact
be the matrix whose $i,k$ element is the $k$-th coordinate of
$v^i$ in some orthonormal basis.  A separable state $R$ (even unnormalized)
has a representation of the form (\ref{eq: decomposition of state})
with $v^i = x^i \otimes y^i$, where we may take $||y^i||=1$ without
loss of generality.  Its eigenvalues are therefore those of the 
Gram matrix $G$ with elements
\beq
G_{ij} = x^{i\dagger} x^j y^{i\dagger} y^j \;.
\eeq
The marginal state on the first factor is 
$\sum_i x^i x^{i\dagger}$, whose eigenvalues are those of $H$
whose elements are
\beq
H_{ij} = x^{i\dagger} x^j\;.
\eeq
But
\beq
G = B \circ H\;,
\eeq
where $B$ is the Hermitian PSD matrix, with ones on the diagonal,
whose elements are $B_{ij} = y^{i\dagger} y^j$.  Therefore
(by the abovementioned fact that the eigenvalues of the
output of a doubly  stochastic
map applied to a Hermitian operator are majorized by those of the Hermitian
input), 
$G$'s eigenvalues are majorized by $H$'s, proving the statement. 
\QED

\section{Comparison with an approach via John's theorem}

A celebrated  result of Fritz John \cite{J48} 
is a natural tool for approaching this problem, so we verify here
that our methods provide stronger results than one can get by 
straightforward application of John's theorem.  
John's theorem gives a shrinking factor such that,
when the smallest ellipsoid covering a convex set is shrunk by that factor,
it fits inside the set.  This is interesting in itself; and if we know 
the ellipsoid, then we can obtain (from its shortest axis) a ball that
fits inside the set as well.

\subsection{The inner and outer ellipsoids, the coefficient of symmetry, 
and John's theorem}

Let $S$ be a closed compact convex set (of nonzero measure, i.e. generating the
vector space $V$) in a real vector space $V$ of dimension $D$.
Let $E_{\rm out}$ be the least-volume 
ellipsoid containing $S$.  Let $S_{\rm centered}$ be $S$ translated
so that the center of its $E_{\rm out}$ is at the origin.  Define 
the ``coefficient of symmetry'' of $S$ as the largest 
``shrinking factor'' $0 \le \alpha \le 1$ such that for 
every $x$ in $S_{\rm centered}$,   $- \alpha x$ is also in 
$S_{\rm centered}$.    John's result
states that if we shrink the least-volume covering ellipsoid, $E_{\rm out}$, 
by multiplying it by a factor $\sqrt{\alpha/{D}}$ the resulting shrunken
ellipsoid is contained in $S$.  Note that when a set $S$ is symmetric
under the action of a compact group $G$, so are $E_{\rm in}(S)$ and 
$E_{\rm out}(S)$.  

\subsection{Application of John's theorem to the set of normalized 
separable states}

Every ellipsoid in the normalized quantum states
is a set of the form:
$L_Q := \{\rho: Q(\rho - I/d, \rho - I/d) \le 1\}$, for a quadratic
form $Q$ that is 
strictly positive semidefinite on the positive semidefinite
matrice, and for every such form $Q$, $L_Q$ is an ellipsoid.  

\begin{proposition}
Let $E = \{\rho: Q_{min}(\rho - I/d, \rho - I/d) \le 1 \}$ be the 
minimum-volume ellipsoid covering the $m$-partite separable 
normalized density matrices.   Let $d = \pi_{i=1}^m d_i$, 
where $d_i$ are the local dimensions.  Then 
\beq \label{eq: the proposition}
\{\rho: Q_{min}(\rho - I/d, \rho - I/d) 
\le \frac{1}{d^{2}(d-1)} \} =: \frac{E}{d\sqrt{d-1}} \subset S\;.
\eeq   
\end{proposition}

\noindent
{\bf Proof:}   We first calculate the  
coefficient of symmetry, by noting
that the quantification over $x$ in the definition of $\alpha$ can 
be restricted to $x$ extreme in $S_{\rm centered} := S - I/d$, i.e. shifted
versions $\pi - I/d$ of pure separable states $\pi$.
Let $x$ be an arbitrary extremal state; 
we find the largest $\alpha$ such that $- \alpha x \in S_{\rm centered}$. 
That is, we seek the largest $\alpha$ such that 
\beqa \label{shooby dooby doo wah day}
- \alpha (\pi - I/d) \in S - I/d, {\rm ~i.e.} \nonumber \\
(1 + \alpha) I/d - \alpha \pi \in S\;.
\eeqa
The LHS of (\ref{shooby dooby doo wah day}) (which we'll  call 
$\Lambda_\alpha$)
has unit trace for all $\alpha$, and is 
PSD (certainly a necessary condition for its separability)
as long as $\alpha < 1/(d-1)$.  With this value of $\alpha$, 
it becomes:
\beq   \label{snoopy}
\Lambda := \frac{1}{d-1} (I - \pi)\;.
\eeq 
We now show that $\Lambda$ is separable, so the coefficient of
symmetry is $1/(d-1)$.
Since $\pi$ is separable, it is equal to $x^{(1)} x^{(1)\dagger} 
\otimes x^{(2)} x^{(2)\dagger}  \otimes \cdots \otimes 
x^{(m)} x^{(m)\dagger} $ for some normalized vectors $x^{(m)} \in H_m$. 
For each $p \in \{1,...,m\}$ 
let $\{x^{(p)}_i\}_{i \in \{1,...,d_p\}}$
be a complete orthonormal basis with first member $x^{(p)}$.
Then, since $\sum_{i_1,i_2,...,i_m} x^{(1)}_{i_1} x^{(1)\dagger}_{i_1} 
\otimes
\cdots \otimes 
x_{i_m}^{(m)} x_{i_m}^{(m)\dagger}= I$, (\ref{snoopy}) becomes
\beq
\sum_{(i_1, i_2, ..., i_m ) \ne (1,1,...,1)} 
\frac{1}{d-1}
x^{(1)}_{i_1} x^{(1)\dagger}_{i_1} \otimes
\cdots \otimes x_{i_m}^{(m)} x_{i_m}^{(m)\dagger}\;.
\eeq
This expresses $\Lambda$ as a convex combination of separable
pure states, demonstrating $\Lambda$'s separability.

Since $\alpha = 1/(d-1)$ and
$D = d^2$ (the 
dimension of the real linear space 
$\ch(d)$ of $d \times d$ Hermitian matrices)
 we have
$\sqrt{\alpha/D} = (1/d)\sqrt{1/(d-1)}$.  
John's theorem then gives (\ref{eq: the proposition}).
\QED

\noindent{\bf Remark: } 
The smallest ball $B$ covering $S$ is centered at $I/d$ and has
radius $\sqrt{(d-1)/d}$.  This  
follows from the easy fact that the pure separable states (indeed
all pure states) lie on the boundary
of this ball, which by unitary invariance therefore contains
{\em all} the normalized states, including the separable ones.   
If this ball were $E$ then (\ref{eq: the proposition})
would give us a ball of radius $O(d^{-3/2})$ inside the Hermitian
matrices.  When the system consists of $m$ $d_0$-dimensional systems,
this is $O(d_0^{-(3/2)m})$.  For qubits, this would have the same
exponent as the results in \cite{GB03}, though it would still be less
good for general $d_0$ (where \cite{GB03} gives
$2^{-(1/2)m}d_0^{-m}$).  The results we obtain elsewhere in this
paper always have a better exponent, though it converges to the
exponent of our earlier result as dimension grows.  $E$ is in fact not
a ball (we thank Stanislaw Szarek for pointing this out to us).
Still, the above result establishes that straightforward application
of John's theorem does not give us better results than \cite{GB03} or
the techniques we use in the other sections of the present paper.  
The largest ball we can
straightforwardly get via John's theorem is the largest ball in the
shrunken minimum-volume ellipsoid, whose radius is $1/(d\sqrt{d-1})$ times the
length of the least principal axis of the covering ellipsoid $E$.  This
must be no larger than $d^{-3/2} \equiv (1/(d\sqrt{d-1})) \sqrt{(d-1)/d}$,
for if the least principal axis of $E$ were larger than the radius
$\sqrt{(d-1)/d}$ of the smallest covering ball $B$ then 
$E$ could not be minimum-volume.

We note that a natural approach to obtaining
$E$ itself is to use some of the
more elementary aspects of the methods exposed in 
\cite{BarBlekh03}:  noting that $E_{\rm out} =:E$ (and $E_{\rm in}$)
must be invariant under the action of conjugation by 
local unitaries $U_1 \otimes U_2 \otimes \cdots \otimes U_m$, 
$E$ must be a ball when restricted to each irrep of this action;
finding the radii of each of these balls determines $E$.

We also note that in the bipartite case, our maximum ball $Ball(r_d)$
of the radius $r_d = 1/\sqrt{d(d-1)}$ in the Frobenius norm is, in
fact, also the maximum-volume ellipsoid inscribed in $S$.  Indeed, it
had been proved in \cite{GB02} that this ball $Ball(r_d)$ belongs to
the convex compact set of normalized separable bipartite states; on
the other hand it is easy to show that $Ball(r_d)$ is the
maximum-volume ellipsoid inside the (larger) convex compact set of all
normalized bipartite states.

\noindent
{\bf Remark.}  Group symmetry can easily be used to compute the coefficient
of symmetry for other convex hulls of orbits of interest in quantum
information theory (and thus when $E_{\rm out}$  can be computed,
one gets
lower estimates of the inner ball's radius via John's theorem).  For example:
\begin{proposition}
Let $F$ 
denote the convex hull of all normalized
``maximally 
entangled states'' of a bipartite system with local dimensions
$n$ (overall dimension $d= n^2$), {\it i.e.} the convex hull of 
the orbit of the state
$\pi := \Psi \Psi^\dagger \in \cb(n,n)$, where
\beq
\Psi = (1/\sqrt{n}) \sum_i e_i \otimes e_i \;,
\eeq
under the action of $U(n) \times U(n)$ on $\cb(n,n)$
as conjugation by local unitaries: $(u,v) \in U(n) \times U(n)$
acts as: $X \mapsto (U(u) \otimes V(v)) X (U(u)^\dagger \otimes V(v)^\dagger)$,
$U,V$ being standard $n\times n$ matrix representations.  
The coefficient of symmetry of $F$ is $1/(d - 1)$.
\end{proposition}

\noindent {\bf Proof:}  As before, by symmetry it suffices to 
find, for a single extremal state $\pi \in F$ (for which 
we choose $\pi$ as defined in the Proposition), the largest
$\alpha$ such that (\ref{shooby dooby doo wah day}) holds, 
with $F$ substituted for $S$.  Exactly as before, we get
$\alpha \le 1/(d-1) \equiv 1/(n^2-1)$ 
necessary for positivity.  We want to 
show that when $\alpha = 1/(n^2-1)$, the state 
\beqa
(1 + \alpha)I/d
- \alpha \pi = (1/(n^2-1))(I - \pi)=:R
\eeqa 
is not only positive but in $F$.  
To this end we use the Choi/Jamiolkowski
isomorphism, and view the matrix $R$ as associated with a map $T$.
$I$ is the Choi matrix of the map $Z: X \mapsto 
(\tr X) I$, i.e. the projector onto the one-dimensional
subspace of matrices 
spanned by the identity,  while $\pi$ is the Choi matrix of $1/n$
times the identity map $\id: X \mapsto X$.  
Therefore $R$ is the Choi matrix of 
\beq
\frac{1}{n^2-1} (Z - \frac{1}{n} \id)
\eeq
For every finite dimension $n$, there exists at least one orthogonal
basis $U_i, i \in {0,...,{n^2-1}}$ for $M(n)$ with $U_0 := I$, and
all $U_i$ unitary.  (For example consider the basis 
$\{P^k S^l : k,l \in \{0,...,n-1\}\}$, with $P$ the diagonal
matrix whose $j$-th diagonal element is $\omega^{j-1}$ for some
primitive $n$-th root of unity $\omega$, and $S$ is the matrix
with elements $S_{ij} = \delta_{(i+1){\rm ~mod~}n,j{\rm~mod~}n}$;
the general question of which such bases exist is considered in 
\cite{Werner2001b}.)   
It is easily verified (cf. e.g. \cite{Werner2001b})
that for any such basis 
the map $Z$ may be written
\beq
Z: X \mapsto (1/n) \sum_{i=0}^{n^2-1} U_i X U_i^\dagger\;.
\eeq
Therefore, with the notation $T_A$ for the map
$X \mapsto AXA^\dagger$,  $R$ is the Choi matrix of
\beq \label{snooker}
\frac{1}{n(n^2-1)} \sum_{i=1}^{n^2-1} T_{U_i}\;.
\eeq 
Since $T_A$ has Choi matrix 
$n(I \otimes A) \pi (I \otimes A^\dagger)$, 
(\ref{snooker}) implies
\beq
R = \frac{1}{n^2-1}
\sum_{i=1}^{n^2-1} (I \otimes U_i) \pi (I \otimes U_i^\dagger)\;,
\eeq
which expresses it as a convex combination of local unitary 
transforms of $\pi$, as desired.  \QED

\section{Application to thermal NMR states and pseudopure states}

In many interesting experimental or theoretical situations, the system
is in a ``pseudopure state'': a mixture of the uniform density matrix
with some pure state $\pi$: \beqa \label{pseudopure state}
\rho_{\epsilon, \pi} := \epsilon \pi + (1 - \epsilon) I/d\;, \eeqa
where $d = d_1,..d_m$ is the total dimension of the system.  For
example, consider nuclear magnetic resonance
(NMR) quantum information-processing (QIP), where $d=2$
(the Hilbert space of a nuclear spin), and $m$ is the number of
spins addressed in the molecule being used.  As discussed in
more detail below, the initialization procedures standard in most
NMRQIP implementations prepare pseudopure states.  

Using Corollary \ref{corollary: general normalized}, with 
$b$  a lower bound on the unnormalized
$2$-norm ball around $I$, $\rho_{\epsilon, \pi}$ 
is separable if 
\beqa
\epsilon \le (b/d)  \sqrt{\frac{d-1}{d-b^2}} \le b/\sqrt{d(d-1)}\;,
\eeqa
For $m$ $d_0$-dimensional systems (so $d={d_0}^m$), this implies
the (negligibly loosened) bound 
\beq \label{eq: earlier bound}
\epsilon \le b/{d_0}^m \;.
\eeq
Since we have established in this paper a bound of 
$b = (d_0/(2d_0 - 1))^{m/2-1}$, we obtain
\beq
\epsilon \le \frac{1}{d_0^{m/2+1} (2 d_0 - 1)^{m/2-1}}\;,
\eeq
This is an exponential improvement over the result in \cite{Rungta2001a}
(the qubit case is in \cite{Braunstein99a})
of $\epsilon \le 1/(1 + d_0^{2m-1})$, and indeed over our results
in \cite{GB03}, although as $d \rightarrow
\infty$ the improvement in the exponent of $m$ over that in \cite{GB03}
goes to zero.

In
liquid-state NMR at high temperature $T$, 
the sample is placed in a high DC magnetic field of strength $B$.
Each spin is in a thermal mixed state, with probabilities
for its two states (aligned ($\uparrow$)  or anti-aligned 
($\downarrow$) with the field) 
proportional to $e^{\pm \beta \mu B}$, where $\beta \equiv 1/kT$ with 
$k$ Boltzmann's constant, $\mu$ the magnetic moment of the nuclear
spin.  For realistic high-T liquid
NMR values of $T=300$ Kelvin, $B = 11$ Tesla, $\eta := \beta \mu B \approx 
3.746 \times 10^{-5} \ll 1$.  Since
$e^{\pm \eta} \approx 1 \pm \eta$, the probabilities
are $p_{\uparrow} 
\approx (1 - \eta)/2$, 
$p_{\downarrow} 
\approx (1 + \eta)/2$.
Thus the thermal density matrix is approximately 
\beqa \label{thermal density matrix}
 \rho = \left( \begin{array}{cc}
		  \frac{1 + \eta}{2} & 0 \\
		  0 & \frac{1- \eta}{2}  
\end{array} \right)^{\otimes n}
\eeqa
(with each qubit expressed 
in the $|\smash\uparrow\rangle, |\smash\downarrow\rangle$ basis).
The highest-probability pure state of independent distinguishable 
nuclear spins, 
has all $m$ spins up and probability about 
$(1 + \eta)^m/2^m \approx (1 + m \eta)/2^m$.
Standard pseudopure-state preparation creates a mixture 
\beqa
(1 - \epsilon) I/2^m + \epsilon \proj{\uparrow \cdots \uparrow}\;,
\eeqa
where
\beqa \label{pseudopure polarization}
\epsilon = \eta m/2^m\;.
\eeqa
of this most
probable pure state
and the maximally mixed state, by applying a randomly chosen
unitary 
from the group of unitaries fixing the all-spins-aligned state.
With 
$\eta \approx 3.746 \times 10^{-5}$,
this implies that below  36 qubits, 
NMR pseudopure states are all separable, compared to 
the $\approx 23$ qubits obtained in \cite{GB03},
and the $\approx 13$ qubits one gets from the bound in 
\cite{Braunstein99a}.
Since we have not shown that the bounds herein are tight, with 
our assumed $\eta$ even
at 36 qubits there is no guarantee one can prepare an entangled 
pseudopure state by randomization.  
We remind the reader, also, that if such a state
existed, there would still be no way of partitioning the qubits
so that the state exhibited bipartite entanglement; as noted in 
\cite{GB03}, the results of \cite{GB02} imply that for the parameters
used above,  
one needs $m = 1/\eta$ qubits (about $26,700$ for our $\eta$) 
before the  pseudopure state obtained from the thermal state by the
randomization procedure described above fails to satisfy \cite{GB02}'s
sufficient
criteria for 
bipartite separability with respect to any
partition of the qubits into two sets.

Schulman and Vazirani's algorithmic cooling protocol \cite{SV99}
shows that it is, in theory, possible to prepare any entangled
state of $n$ qubits from polynomially many (in $n$) thermal NMR qubits,
although the overhead is discouraging.  
The question of just how many qubits are required by means 
possibly simpler than algorithmic cooling is also of interest.
One can gain some information about this using our results, by
applying Corollary \ref{corollary: general normalized}
 to the initial thermal density matrix of an NMR 
system.  For the initial thermal density matrix $\rho$ of 
(\ref{thermal density matrix}),we have:
\beqa
||\rho - I/d ||_2^2 = \frac{(1 + \eta^2)^m -1}{2^m} 
\approx m \eta^2/2^m\;.
\eeqa
This should be compared to the separability condition obtained
by using the relation (\ref{eq: better recursion relation}) for $a_n$, and
Corolary \ref{corollary: general normalized}.    
Numerical comparison shows that 
17 qubits are required before this bound is exceeded (rather than
the 36 required for the pseudopure state prepared from this thermal
state).  (Our earlier bound allowed only the weaker statement
that for fewer than 14 qubits, no entanglement exists in the
thermal state \cite{GB03}.)

\section*{Acknowledgments}
We thank Adam Sears for help with the numerical comparisons of our bounds
with the thermal and pseudopure NMR states, 
Ike Chuang and Manny Knill for discussions, Stanislaw Szarek for 
enlightenment about his results and about the shape of the minimum-volume
covering ellipsoid, and Roland Hildebrand for informing us about his work.  
We thank the US DOE for financial
support through Los Alamos National Laboratory's Laboratory Directed 
Research and Development (LDRD) program, and ARDA and the NSA for 
support.

\begin{appendix}
\section{Contraction bound for stochastic ball-positive maps on all matrices}

In this section, we state and prove a contraction bound from $2$-norm 
to $\infty$-norm (i.e. a bound on the induced operator norm) for stochastic,
ball-positive maps on {\em all} matrices.  It is slightly more involved to prove
than the one for maps on 
traceless matrices used in the body of the paper, but although we ultimately
did not need it for the present paper, we present it here in the hope
that it may find uses elsewhere in quantum
information theory or mathematics.

\begin{definition}
Define $\lambda'(d_1,d_2, a)$ as the maximum, over all    
stochastic maps $\phi: M(d_2) \rightarrow M(d_1)$, 
positive on $G(d_2,a)$, and over all $X \in M(d_2)$, of  
$||\phi(X)||_\infty/||X||_2$.
\end{definition}

\begin{proposition} \label{prop: lambdaprime bound}
\beq
\lambda'(d_2, d_1, a) = \sqrt{\frac{2}{a^2} - \frac{1}{d_2}}\;.
\eeq
\end{proposition}

\noindent
{\bf Proof:}  
Recall from (\ref{schlemiel}) that for Hermitian traceless
\beqa \label{schlemiel2}
||\phi(\Delta)||_\infty / ||\Delta||_2 \le 
(1/a) \sqrt{1 - a^2/d }\;. \label{groovy2}
\eeqa
To extend this to arbitrary, not necessarily Hermitian traceless,
matrices consider:
\beqa \label{schlemazel2}
M = c(I/\sqrt{d }) + B\;,
\eeqa
with $B$ traceless but not necessarily Hermitian.  To bound
$||\phi(M)||_\infty/||M||_2$ it suffices by homogeneity to bound
it for $||M||_2=1$, i.e. defining $||B||_2 =:b$, for $c^2 + b^2= 1$.
Writing $B$ in terms of Hermitian and antiHermitian parts
as $B =  X + iY$, we have:
\beqa \label{schnuck2}
||\phi(M)||_\infty & \le & c/\sqrt{d } + ||\phi(X)||_\infty + ||\Phi(Y)||_\infty
\nonumber \\
& \le & c/\sqrt{d } + a^{-1}\sqrt{1-a^2/d } (||X||_2 + ||Y||_2) \nonumber \\
& \le & c/\sqrt{d } + a^{-1}\sqrt{1-a^2/d } \sqrt{2} b\;, \label{RHS2}
\eeqa
where the second inequality is by (\ref{groovy2}) and the last is
elementary Euclidean geometry.
Defining
\beqa \label{def: gamma}
\gamma := a^{-1}\sqrt{2(1 - a^2/d)}\;,
\eeqa
we maximize the RHS of (\ref{RHS2})
over $c, b$ such that $c^2+b^2=1$ (i.e. $||M||_2 = 1$).  We obtain
\beqa
c = \sqrt{\frac{1}{1 + \gamma^2 d}}, \nonumber \\
b = \sqrt{ \frac{\gamma^2 d}{1 + \gamma^2 d}}\;,
\eeqa
and hence a maximal value for the RHS of
\beqa \label{eq: themax}
\sqrt{1/d + \gamma^2}\;,
\eeqa

Substituting our definition for $\gamma$ gives 
\beqa
||\phi(M)||_\infty \le \sqrt{2/a^2 - 1/d}\;.
\eeqa 
Thus an upper bound on $||\phi(M)||_\infty / ||M||_2$
for arbitrary $M$ and $G(d_2, a)$-positive stochastic
$\phi$ (which is to say on $\lambda(d_2, d_1, a)$)
is $\sqrt{2/a^2 - 1/d_2}$

For the lower bound portion of the proposition , we exhibit a $G(d,
a)$-positive stochastic map map $\tau$ for which
$||\tau(X)||_\infty/||X||_2 = \sqrt{\frac{2}{a^2} - \frac{1}{d}}.$
We begin by defining a family of stochastic maps parametrized by $\mu
\ge 0$, acting on {\em Hermitian} matrices.  For $N \ge 4, d_1 \ge 2$
we define $\tau$ by specifying $\tau(I)= I$, and:
\begin{eqnarray*}
\tau \left( 
 \begin{array}{cccc}
		  1/\sqrt{2} & 0 & 0 & \cdots \\
		  0 & -1/\sqrt{2} & 0 & \cdots \\
0 & 0 & 0 & \cdots \\
\vdots & \vdots & \vdots & \ddots  
\end{array} \right) = \mu
\left( 
 \begin{array}{rrr}
		  1 & 0 &  \cdots \\
		  0 & -1 &  \cdots \\
                  \vdots & \vdots & \ddots      
\end{array} \right) =: \mu \sigma_z \;, \nonumber \\
\tau \left( 
 \begin{array}{ccccc}
		  0 & 0 & 0 & 0 & \cdots \\
                  0 & 0 & 0 & 0 & \cdots \\
		  0 & 0 & -1/\sqrt{2} & 0 & \cdots \\
0 & 0 & 0 & 1/\sqrt{2} & \cdots \\
\vdots & \vdots & \vdots & \vdots & \ddots  
\end{array} \right) = \mu
\left( 
 \begin{array}{ccc}
		  0 & 1 & \dots \\
		  1 & 0 & \dots \\
                  \vdots & \vdots & \ddots
\end{array} \right) =: \mu \sigma_x\;.
\end{eqnarray*}
Dots indicate the matrices are to be filled out with zeros.  
$I$ and the two input matrices given above are mutually orthogonal
in trace inner product;  on the orthocomplement of their span, 
$\tau$ is taken to map everything
to zero.  Call the input matrices above $Z$ and $X$ (so that 
$\tau_\mu(Z) = \sigma_z, \tau_\mu(X) = \sigma_x$). 
$\tau_\mu$ extends to antiHermitian matrices homogeneously, due to its 
Hermiticity
preserving property, so that
$\tau_\mu (iZ) = i\mu \sigma_z$, $\tau_\mu(iX) = i \mu \sigma_x$.  
(The names $\sigma_x, \sigma_z$ are chosen for the output
matrices because the usual Pauli matrices that go by these names
appear in the upper left-hand $2 \times 2$ blocks of our $\sigma_x, 
\sigma_z$, and are padded out with zeros.)

For Hermitian traceless $B$, the maximal
value of $||\tau_\mu(B)||_\infty/||B||_2$ will occur where $B = c Z + b X$.
Then $\tau_\mu(B) = \mu( c \sigma_z + b \sigma_x) \equiv 
\mu(\sqrt{c^2 + b^2}) \sigma_\alpha$, where $\sigma_\alpha$ is some
matrix which has a $2 \times 2$ Hermitian upper left diagonal block with 
eigenvalues $\pm 1$, and is zero elsewhere.  Hence $||\tau(B)||_\infty 
= \mu \sqrt{c^2 + b^2}$, and since $||B||_2= \sqrt{c^2 + b^2}$, 
$||\tau(B)||_\infty/||B||_2 = \mu$.  
However, for $\tau_\mu$ to be $G(d,a)$-positive
requires that  
\beqa \label{schnauzer}
||\tau_\mu(Y)||_\infty / ||Y||_2 \le a^{-1} \sqrt{1 - a^2/d}
\eeqa
hold for all traceless Hermitian $Y$
(cf. (\ref{schlemiel})), so we must have
\beq
\mu \le a^{-1} \sqrt{1 - a^2/d}\;.
\eeq
We choose $\mu$ equal to the RHS here;  then the inequality (\ref{schnauzer})
holds for all Hermitian $Y$, as required for $G(d,a)$-positivity.

Now, we consider  the not-necessarily-traceless
matrix  
$Y = \frac{\alpha}{\sqrt{d}} I + (\beta/\sqrt{2})(X + iZ)$.
Then $||Y||_2 = \sqrt{\alpha^2 + \beta^2}$, which we set equal to one
WLOG.  Now,
\beqa
||\tau(Y)||_\infty = \alpha/\sqrt{d} + (\beta/\sqrt{2})
|| \phi(X) + i \phi(Z)||_\infty \nonumber \\
= \alpha/ \sqrt{d} + \beta \mu ||\sigma_x + i \sigma_z||_\infty \nonumber \\
= \alpha/\sqrt{d} + \beta a^{-1} \sqrt{1 - a^2/d}\sqrt{2}\;.
\eeqa
The last equality uses just the definition of $\mu$ and the
result $||\sigma_x + i \sigma_z||_\infty = 2$.  The latter is easily 
obtained by  noting that 
\beq
\left( \begin{array}{cc}                      
                  i & 1 \\
		  1 & -i \end{array} \right)
\left( \begin{array}{c}                      
                  1/\sqrt{2} \\
		  i/\sqrt{2} \end{array} \right) = 
\left( \begin{array}{c}                      
                  i \sqrt{2} \\
		  \sqrt{2} \end{array} \right)\;.
\eeq
This vector $[i \sqrt2, \sqrt{2}]^t$ has Euclidean norm $2$, which is
therefore a lower bound on the operator norm of $\sigma_x + i \sigma_y$;
since the Frobenius norm upper-bounds the operator norm, and is equal
to $2$ in this case, the operator norm is 2.  
Define 
\beq
\gamma' := a^{-1} \sqrt{2(1 - a^2/d)} \equiv \sqrt{2/a^2 - 2/d}\;.
\eeq
Then we have  
\beq
||\tau(Y)||_\infty = \alpha/\sqrt{d} + \beta \gamma'\;
\eeq
and the same argument used to obtain 
(\ref{eq: themax}) as the maximum of (\ref{RHS}) yields 
$\sqrt{1/d + \gamma^{'2}}$ as the maximum here.  Substituting
the definition of $\gamma'$ gives 
a maximum of $\sqrt{2/a^2 - 1/d}$
for $\lambda$, which matches the previously obtained upper bound.
\QED

\end{appendix}


\end{document}